\documentclass[12pt, reqno] {article}
\usepackage{latexsym}
\usepackage{amsmath}
\usepackage{mathrsfs}
\usepackage{amsfonts}
\usepackage{graphicx}
\usepackage{wrapfig}
\usepackage{xcolor}
\usepackage[totalheight = 21cm, totalwidth = 17cm]{geometry}
\usepackage{aas_macros}

\newcommand{\bea}{\begin{eqnarray}}
\newcommand{\eea}{\end{eqnarray}}
\newcommand{\be}{\begin{equation}}
\newcommand{\ee}{\end{equation}}
\numberwithin{equation}{section}

\allowdisplaybreaks

\begin{document}
\begin{titlepage}  
\pagestyle{empty}
\baselineskip=21pt
\vspace{1cm}
\begin{center}
{\it Awarded Fourth Prize in the 2024 Gravity Research Foundation \\ Competition for Essays on Gravitation}
\end{center}
\vspace{1cm}
\begin{center}
{\bf {\large In Search of the Biggest Bangs since the Big Bang}}
\end{center}
\begin{center}
\vskip 0.3in
{\bf John~Ellis}$^{1,2,*}$, {\bf Malcolm~Fairbairn}$^{1}$, {\bf Juan~Urrutia}$^{3,4}$
and {\bf Ville Vaskonen}$^{4,5,6}$
\vskip 0.1in
{\it
$^1${TPPC Group, Physics Department, 
King's College London, Strand WC2R 2LS, UK}\\
$^2${Theoretical Physics Department, CERN, CH-1211 Geneva 23, Switzerland}\\
$^3${Keemilise ja Bioloogilise F\"u\"usika Instituut, R\"avala pst. 10, 10143 Tallinn, Estonia}\\
$^4${Department of Cybernetics, Tallinn University of Technology, Akadeemia tee 21, 12618 Tallinn, Estonia}\\
$^5${Dipartimento di Fisica e Astronomia, Universit\`a degli Studi di Padova, Via Marzolo 8, 35131 Padova, Italy}\\
$^6${Istituto Nazionale di Fisica Nucleare, Sezione di Padova, Via Marzolo 8, 35131 Padova, Italy}} 

\vspace{1cm}
{\bf Abstract}
\end{center}
\baselineskip=18pt \noindent  

Many galaxies contain supermassive black holes (SMBHs), whose formation and history raise many puzzles. Pulsar timing arrays have recently discovered a low-frequency cosmological ``hum" of gravitational waves that may be emitted by SMBH binary systems, and the JWST and other telescopes have discovered an unexpectedly large  population of high-redshift SMBHs. We argue that these two discoveries may be linked, and that they may enhance the prospects for measuring gravitational waves emitted during the mergers of massive black holes, thereby opening the way towards resolving many puzzles about SMBHs as well as providing new opportunities to probe general relativity.\\
~~\\
~~\\

\vspace{0.25cm}

\leftline{KCL-PH-TH/2024-27, CERN-TH-2024-058, AION-REPORT/2024-04}
\end{titlepage}
\baselineskip=18pt

\section{Introduction} 

Supermassive black holes (SMBHs) that weigh
millions, even billions of times more than the Sun lurk at the centres of many galaxies. Forming these compact objects would have required the emission of
unparallelled amounts of binding energy in the form of gravitational waves in the biggest bangs since
the Big Bang. However, until now the formation histories of SMBHs has remained largely an arena for
theoretical modelling and speculation. Were they formed directly in the collapses of gas clouds or the mergers of SMBH-less
protogalaxies? Or via some hierarchical process involving the mergers of
intermediate-mass black holes already present in protogalaxies? Or even more hierarchically via
a tree of successive mergers seeded by black holes formed in the collapses of an early generation of stars?

Two sets of recent measurements bring these questions into sharp focus, leading to hopes that we may
soon observe how SMBHs form and evolve. One set of measurements has been provided by pulsar timing
arrays (PTAs) that have discovered a stochastic gravitational wave background (SGWB) at frequencies in
the nHz range~\cite{NANOGrav:2023gor,Reardon:2023gzh,Xu:2023wog,EPTA:2023xxk}, whose default astrophysical interpretation invokes SMBH binary systems as sources. The
other set of measurements has been of a population of high-redshift SMBHs by the JWST~\cite{2023A&A...677A.145U,CEERSTeam:2023qgy,2023ApJ...959...39H,bogdan2023evidence,Ding_2023,Maiolino:2023bpi,yue2023eiger} and other
telescopes (see, e.g.,~\cite{2021ApJ...914...36I}). The strengths of both of these signals challenged previous astrophysical expectations.
We argue that these discoveries may be linked~\cite{Ellis:2024wdh}, and that they may suggest a higher abundance of low-redshift SMBHs
than had been expected previously, enhancing the prospects for observing directly the
mergers of massive black holes with upcoming gravitational wave detectors and probing astrophysics and general relativity.

Pulsars emit extremely regular signals whose arrival times could be affected by the passage of gravitational waves.
The PTAs searched for this effect, which would be manifested as a correlated stochastic noise delaying slightly the arrival times of the signals from 
dozens of pulsars. In 1982 Hellings and Downs predicted that timing noise induced by gravitational waves
would exhibit a characteristic angular correlation between pulsars in different directions~\cite{1983ApJ...265L..39H}, and in 2001 Phinney 
predicted that SMBH binary systems would produce such an effect with a characteristic spectral index~\cite{Phinney:2001di}. 
In mid-2023 the PTAs reported a common noise source with the Hellings-Downs angular correlation and
a spectral index similar to Phinney's prediction, evidence for a SGWB that might be due to SMBH binaries. However, the strength of the SGWB was larger than anticipated, and the spectral shape indicated did not agree very well with predictions based on gravitational wave emission alone, as we discuss below. 

In parallel, groups using the JWST have reported the discovery of a population of SMBHs with masses 
$\sim 10^6 - 10^9$ solar masses at redshifts $z > 4$~\cite{2023A&A...677A.145U,CEERSTeam:2023qgy,2023ApJ...959...39H,bogdan2023evidence,Ding_2023,Maiolino:2023bpi,yue2023eiger}. These observations are consistent with previous measurements from other telescopes of SMBHs at redshifts $z > 6$~\cite{2021ApJ...914...36I}.
However, the abundance of these high-$z$ SMBHs was surprisingly large, and the observations do not obey the relationship between host galaxy mass-SMBH mass observed in active galactic nuclei (AGNs) at low redshifts~\cite{2015ApJ...813...82R}. This has
triggered suggestions that the JWST sample might have been biased by selection effects~\cite{2024arXiv240300074L}, or that
the SMBH-galaxy mass relation for AGNs might have evolved strongly with redshift~\cite{Pacucci:2024ijt}.  

In this essay we argue that the numbers of high-redshift black holes and the strength of the PTA
gravitational wave noise  can be understood in a consistent framework~\cite{Ellis:2024wdh}. This suggests good prospects~\cite{Ellis:2023dgf} that upcoming
experiments looking for gravitational waves with higher frequencies than those detected by the PTAs
should be able to detect emissions from mergers of black holes much heavier than those detected by the
LIGO, Virgo and KAGRA experiments. Their measurements will provide information about the seeds of the
present SMBH population, revealing how they were formed, and also provide new probes of general relativity. Their measurements could also help distinguish between astrophysical models of the SGWB and cosmological scenarios invoking physics beyond the Standard Model~\cite{Ellis:2023oxs}.

\section{Binary Interpretation of Pulsar Timing Array Data}

\begin{wrapfigure}[13]{lt}{0.5\textwidth}
\vspace{-0.85cm}
\includegraphics[width=0.5\textwidth]{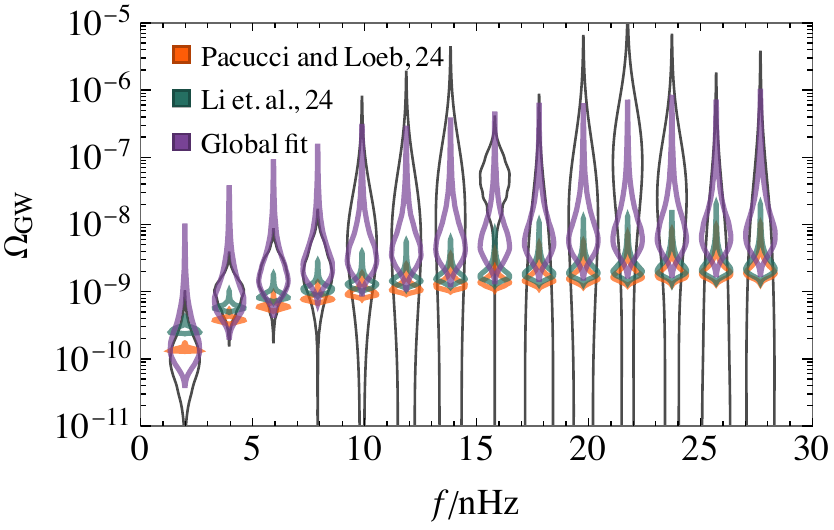}
\vspace{-1.1cm}
\caption{\it NANOGrav data (grey ``violins")~\cite{NANOGrav:2023gor} compared with a SMBH binary model incorporating 
environmental effects (purple)~\cite{Ellis:2023dgf} and two other models with different SMBH-galaxy mass relations (green and orange)~\cite{2024arXiv240300074L,Pacucci:2024ijt}.}
\label{fig:violins}
\end{wrapfigure}
Four pulsar timing array (PTA) collaborations have recently reported data on the SGWB~\cite{NANOGrav:2023gor,Reardon:2023gzh,Xu:2023wog,EPTA:2023xxk}. We focus here on the data from the NANOGrav collaboration shown in Fig.~\ref{fig:violins}, which presented the strongest evidence for a SGWB.  The strength of the signal in each frequency bin is represented as a ``violin" whose varying width corresponds to the posterior probability density. 

Also plotted are the predictions of three models (in orange, green and purple) that attempt to fit the data using a population of SMBH binaries at high redshift.  
These models make different assumptions about the ratios of the masses of SMBHs and their host galaxies. Two of the models predict SGWB signals that are weaker than the NANOGrav measurements~\cite{2024arXiv240300074L,Pacucci:2024ijt}, though they are compatible at the 3$\sigma$ level with the frequency dependence expected by SMBH binaries losing energy via gravitational wave emission. However, the bins at lower frequencies, corresponding to larger binary separations, suggest that other energy-loss mechanisms may also be at work, presumably interactions with the galactic environment. The purple ``violins" have been calculated on the basis of the global fit to SMBH data discussed below and allowing for possible environmental effects~\cite{Ellis:2023dgf}: they fit the data well.

\section{JWST vs PTAs}
\begin{wrapfigure}[20]{lt}{0.5\textwidth}
\vspace{-0.5cm}
\includegraphics[width=0.47\textwidth]{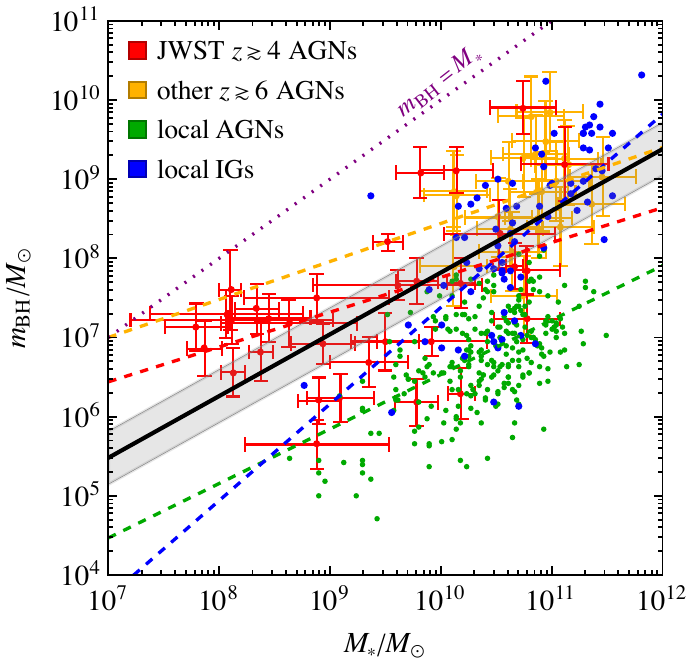}
\vspace{-0.5cm}
\caption{\it Measurements of SMBH masses as functions of the stellar masses in galaxies, including JWST~\cite{2023A&A...677A.145U,CEERSTeam:2023qgy,2023ApJ...959...39H,bogdan2023evidence,Ding_2023,Maiolino:2023bpi,yue2023eiger} and other high-redshift measurements~\cite{2021ApJ...914...36I},
as well as low-redshift (local) AGNs and inactive galaxies (IGs)~\cite{2015ApJ...813...82R}. The black line and band are from a global fit omitting the low-redshift AGNs~\cite{Ellis:2024wdh}.}
\label{fig:massrelations}
\end{wrapfigure}

As already noted, one of the most surprising early observations with the JWST has been the discovery of a population of high-redshift SMBHs, shown in
red in Fig.~\ref{fig:massrelations}~\cite{2023A&A...677A.145U,CEERSTeam:2023qgy,2023ApJ...959...39H,bogdan2023evidence,Ding_2023,Maiolino:2023bpi,yue2023eiger}. 
This population looks very different from the SMBHs in low-redshift AGNs (shown in green), though it is
less incompatible with the population in low-redshift inactive galaxies (IGs) (shown in blue)~\cite{2015ApJ...813...82R}, and is also compatible with the trend of previous
observations of high-redshift AGNs (shown in yellow)~\cite{2021ApJ...914...36I}. A global fit to the high-redshift and inactive galaxy data~\cite{Ellis:2024wdh} is shown as the black band in
Fig.~\ref{fig:massrelations}.

At low redshifts the populations of SMBHs measured in inactive galacties and  in AGNs are significantly different, the latter being much lighter, as seen in Fig.~\ref{fig:massrelations}. The reason is not well understood, but it appears that heavier SMBHs are more likely to be inactive while the lighter ones are AGNs. What our global fit shows is that all observations point to approximately the same scaling relation except for the local AGNs. The SMBHs observed with JWST are of similar stellar masses to the local inactive galaxies, as found in our global fit.

This global fit was used to calculate the SMBH binary model results shown in purple in Fig.~\ref{fig:violins},
which are very consistent with the NANOGrav measurements of the SGWB. The JWST and other high-redshift SMBH data seem to telling a similar story to
the PTA data: maybe the JWST results should not have been so surprising?

At least two other explanations why the SMBHs measured by JWST are so heavy have been proposed. One is that the JWST SMBH observations are subject to election bias and measurement errors~\cite{2024arXiv240300074L}, and that the underlying population is actually quite similar to the local AGN population. The other is that the SMBH-stellar mass relation depends strongly on redshift~\cite{Pacucci:2024ijt}, in such a way that the two populations can be reconciled. These two assumptions can explain, under different assumptions, both the local AGNs and the new high-redshift observations, but cannot be reconciled easily with the NANOGrav SGWB measurement.

Fig.~\ref{fig:violins} shows predictions for NANOGrav derived from these two alternative interpretations of the JWST data. The green points are calculated on the hypothesis that the JWST
data were subject to selection bias and measurement errors~\cite{2024arXiv240300074L}, and the orange points are based
on the proposal that the SMBH-stellar mass relation evolved rapidly with redshift~\cite{Pacucci:2024ijt}. We see that both of these models undershoot the NANOGrav. The NANOGrav data on the SGWB seem to confirm that the Universe has long contained more SMBHs than expected from the population of
\begin{wrapfigure}[14]{lt}{0.6\textwidth}
\vspace{-0.4cm}
\includegraphics[width=0.59\textwidth]{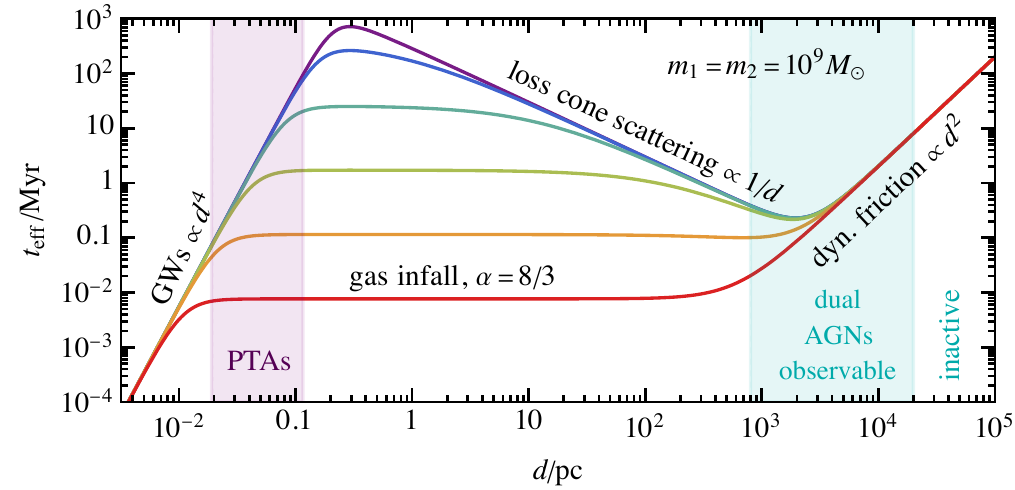}
\caption{\it The high-redshift dual AGNs observed with JWST~\cite{2023arXiv231003067P} and the abundance of little red dots~\cite{2024arXiv240403576K} are consistent with a model of environmental effects that
also fits NANOGrav data~\cite{Ellis:2024wdh}, where details are described. The red and orange curves fit the data best.}
\label{fig:environment}
\end{wrapfigure}
low-redshift AGNs - NANOGrav seem to be detecting a population of SMBHs in active galaxies at high redshift that is similar to the quiescent population today. This suggestion that there may be more low-redshift SMBHs than those in AGNs is good news for the experiments that will be searching for gravitational waves at higher frequencies than the PTAs.

More surprising JWST results include the observation of a high fraction of dual AGNs separated by a few kiloparsecs~\cite{2023arXiv231003067P}, some 20 - 30\% in a sample of 12 AGNs, orders of magnitude higher than previous theoretical estimates and the discovery of the little red dots~\cite{2024arXiv240403576K} in the redshift range $2 < z < 11$ which also have surprised by their high number density. Interpreting AGNs as triggered during mergers, we use this as an observation of the total AGN number.

This fraction and high number density may also be understood 
in a model of environmental effects on the evolution of galaxy pairs and SMBH binaries~\cite{Ellis:2024wdh}. The latter evolves at small separations by emitting gravitational waves, but the low-frequency downturn in the NANOGrav data in Fig.~\ref{fig:violins} seems to require environmental effects~\cite{Ellis:2023dgf}. Several such effects
are expected at larger separations, such as gas infall, viscous drag, stellar loss-cone scattering and dynamical friction. Fig.~\ref{fig:environment} illustrates a model of these effects~\cite{Ellis:2024wdh}. {The red and orange curves fit best the dual AGN data and the number of observed LRDs, assuming that the AGN activity is triggered during the merger, whereas for the other evolutions the dual AGN fraction would be much smaller than seen by JWST. The red and orange curves correspond also to the models that fit best the NANOGrav data~\cite{Ellis:2023dgf}. Maybe the JWST, LRD and dual AGN observations should also not have been so surprising? Moreover, as seen in Fig.~\ref{fig:environment}, dual AGNs would have evolved rapidly into closer binaries which is also potential good news for searches for gravitational waves from SMBH binaries.}

\section{Prospects for Future Measurements}

\begin{wrapfigure}[13]{lt}{0.5\textwidth}
\vspace{-0.5cm}
\includegraphics[width=0.5\textwidth]{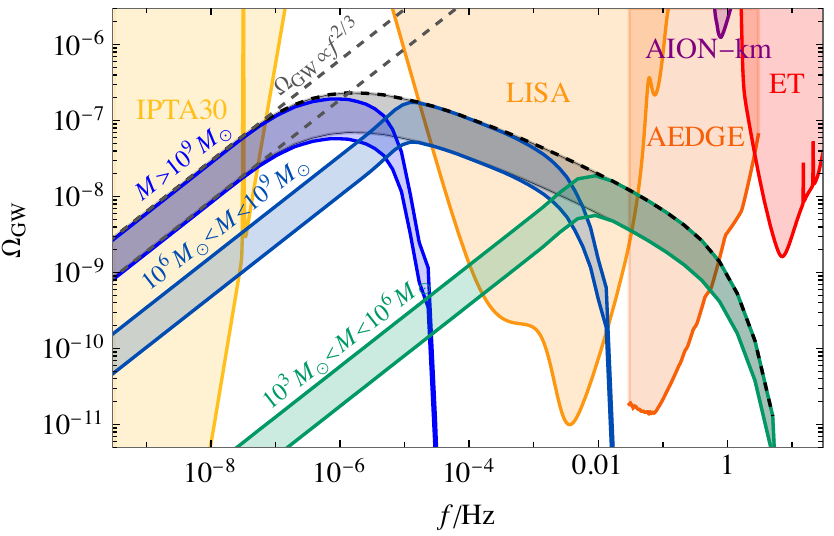}
\vspace{-1.1cm}
\caption{\it Estimates of the potential sensitivities of gravitational wave
experiments to mergers of black holes in different mass ranges~\cite{Ellis:2023dgf,Ellis:2023owy}.}
\label{fig:extrapolation}
\end{wrapfigure}

The PTA experiments are restricted to making measurements in the frequency range $< 30$~nHz,
but may nevertheless be able to detect distinctive signatures of the SMBH binary scenario. The dominant
contributions to the SGWB are expected to come from a limited number of nearby binaries, This could lead
to fluctuations in the intensity in different frequency bins, anisotropies and polarization effects (see, e.g.,~\cite{Ellis:2023owy}) that may be observable in future rounds of PTA data.

Using the Extended Press-Schlechter formalism~\cite{Press:1973iz,Bond:1990iw} to estimate the rates for mergers of galactic haloes, and assuming a
constant probability for a halo merger to yield a black hole merger, one can estimate the prospects that
higher-frequency gravitational wave experiments will observe the mergers of lower-mass black holes~\cite{Ellis:2023dgf,Ellis:2023owy}.
As seen in Fig.~\ref{fig:extrapolation}, there may be good prospects for LISA to observe mergers of
black holes with masses in the range $10^6 - 10^9$ solar masses, and an atom interferometer experiment
such as AEDGE~\cite{AEDGE:2019nxb} may be able to measure mergers of black holes weighing $10^3 - 10^6$ solar masses.
\vspace{-5mm}
\section{How were SMBHs Seeded?}

What could one learn from measuring such mergers? As mentioned in the Introduction, there are many scenarios for the seeding of SMBHs, ranging from stellar relics weighing
${\cal O}(100)$ solar masses to the possibility that they were formed directly in the collapses or mergers of protogalaxies.
Assuming the SMBH binary interpretation of the 
PTA data and their extrapolation to higher-frequency experiments
shown in Fig.~\ref{fig:extrapolation}, could these experiments provide an indication how SMBHs were seeded?

\begin{wrapfigure}[16]{lt}{0.5\textwidth}
\vspace{-0.5cm}
\includegraphics[width=0.49\textwidth]{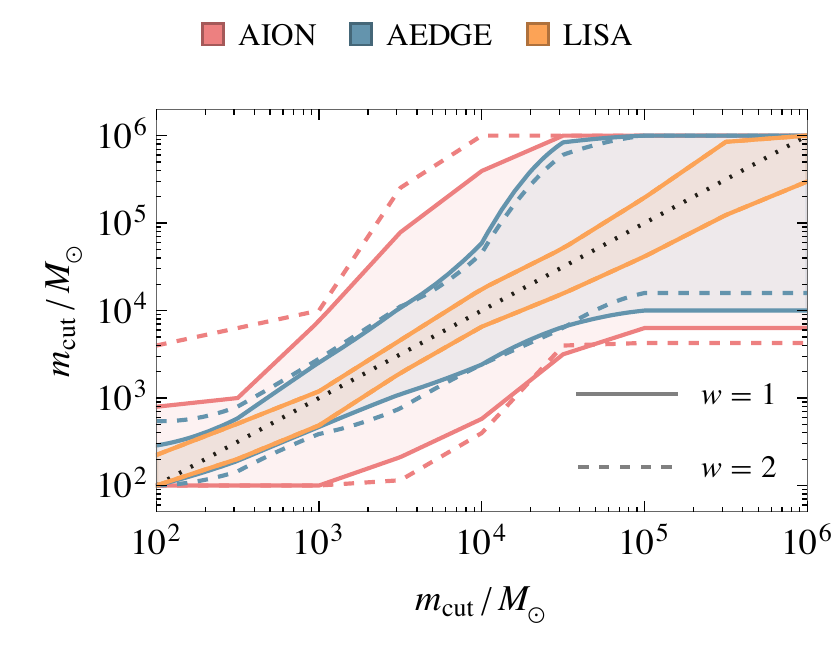}
\vspace{-0.2cm}
\caption{\it The 95\% CL accuracy with which LISA~\cite{Audley:2017drz}, AEDGE~\cite{AEDGE:2019nxb} and AION-1km~\cite{Badurina:2019hst} could measure a nominal cut on the SMBH
seed mass over the range $10^2 - 10^6$ solar masses~\cite{Ellis:2023iyb}.}
\label{fig:seeds}
\end{wrapfigure}

This question is addressed in Fig.~\ref{fig:seeds}~\cite{Ellis:2023iyb}, where we compare the precision with which gravitational wave
measurements by the AION-1km~\cite{Badurina:2019hst}, AEDGE~\cite{AEDGE:2019nxb} and LISA~\cite{Audley:2017drz} experiments could 
determine a lower cut on the seed mass (assumed to have a spread $w = 1$ or 2 on a logarithmic scale) between $10^2 - 10^6$ solar masses: the horizontal axis show the assumed input,
and the vertical axis is the value that could be extracted from the data. Our analysis shows that LISA could make an interesting
measurement of the seed mass over the entire mass range explored, and that AEDGE could make an interesting
measurement if the seed mass is in the range $10^2 - 10^4$ solar masses.

\section{Alternative Interpretations of the PTA Data}

\begin{figure}
\centering
\includegraphics[width=0.8\textwidth]{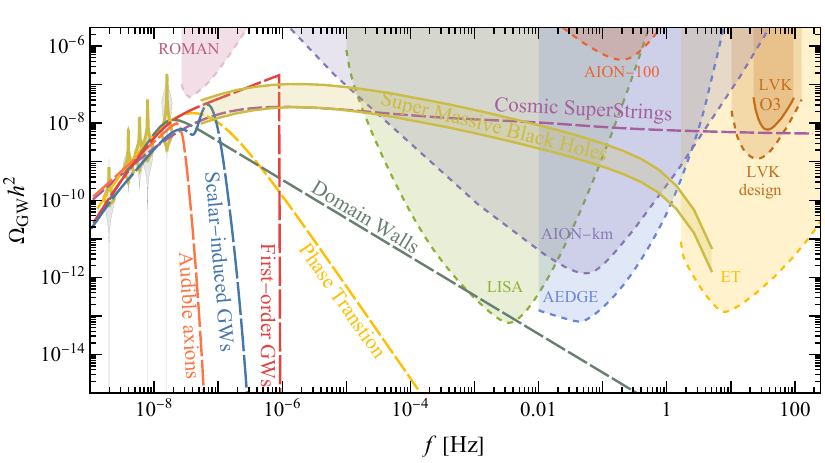}
\caption{\it Compilation of the best fits to the NANOGrav data for SMBH binaries with environmental effects (olive band) and in some cosmological models invoking physics beyond the Standard Model, compared with the sensitivities of different detectors~\cite{Ellis:2023oxs}.}
\label{fig:alternatives}
\end{figure}

Measuring gravitational waves emitted by the mergers of SMBHs - ``the biggest bangs since the Big Bang" - 
would excite astrophysicists, but the default astrophysical interpretation of the PTA data as due to SMBH binaries
is not the only possibility, and cosmologists have made many other proposals that invoke particle physics beyond
the Standard Model. Fig.~\ref{fig:alternatives} compares some of these particle cosmology models with the
NANOGrav data and the SMBH binary model discussed above~\cite{Ellis:2023oxs}.

The list of cosmological models is topped by cosmic strings, lines of energy that may be formed during a phase transition 
in the very early Universe. Subsequently they would move under the influence of gravity and occasionally cross each other, in which case
they might give birth to loops of string that would later collapse through the emission of gravitational waves. This
process would lead to a smooth spectrum extending over a very large range of frequencies. Indeed, this
interpretation of the NANOGrav data is already constrained by data from the LIGO, Virgo and KAGRA experiments,
as seen on the right side of Fig.~\ref{fig:alternatives}, and will be probed further by their future data. 
 
One of the other favoured particle cosmology scenarios listed in Fig.~\ref{fig:alternatives} is a first-order phase transition
that could generate gravitational waves following nucleation and collisions of bubbles of the new vacuum. Interestingly, in order to fit the NANOGrav data
this transition would need to have occurred at an energy scale of a few hundred GeV, within the energy range of the LHC, but in a sector of the theory that is hidden from experiment~\cite{Ellis:2023oxs}. This and the other cosmological scenarios featured
in Fig.~\ref{fig:alternatives} would yield gravitational wave spectra that cut off at frequencies $> 10^{-6}$~Hz,
and so would not be detectable by the higher-frequency gravitational wave experiments shown in Fig.~\ref{fig:extrapolation}~\cite{Ellis:2023oxs}.

Intriguingly, all the cosmological scenarios shown in Fig.~\ref{fig:alternatives} fit the NANOGrav data better (in a Bayesian 
sense) than the default astrophysical SMBH binary scenario, which may excite particle physicists~\cite{Ellis:2023oxs}. However, the preference
is not (yet?) significant: time and more data will tell!

\section{Summary and Prospects}

The discovery of a stochastic background of low-frequency gravitational waves by NANOGrav and other pulsar timing arrays has opened a new
window on the Universe whose source remains to be clarified. The most plausible interpretation of the PTA signal is that it is due to 
supermassive black hole binary systems, though this remains to be confirmed. Possible confirmatory observations could include fluctuations in the 
signal strength between different frequency bins, angular anisotropies in the signal strength, and polarization of the gravitational waves. The black hole binary
interpretation also suggests that higher-frequency gravitational wave detectors such as LISA and AEDGE should be able to observe directly the mergers
of black holes much more massive than those observed by LIGO, Virgo and KAGRA: ``the biggest bangs since the Big Bang". 

We have emphasized in this essay the consistency of this astrophysical interpretation of the PTA data with JWST and other observations of high-redshift
SMBHs, rendering them less surprising than may have seemed initially. We have also discussed the capabilities of gravitational wave measurements to cast light on the mechanisms that contributed to SMBH formation. 

These measurements will also be able to contribute to our understanding of general relativity. Observations of the gravitational waves emitted by massive black hole binaries could extend
over long periods of time - potentially years - making possible detailed inspiral and infall measurements that would yield powerful tests of
high-order calculations of post-Newtonian and post-Minkowskian effects. Also interesting would be measurements of higher-mode contributions to the
frequency spectrum of gravitational wave emissions during the final stages of mergers and the subsequent ringdown phases, which could probe strong gravity beyond the perturbative regime. For these reasons the
astrophysical interpretation of the PTA data is exciting for general relativists as well as astrophysicists.

This said, as also mentioned in this essay, other interpretations of the PTA data are waiting in the wings for the possibility that the astrophysical interpretation fails to be 
confirmed. These include many non-standard cosmological scenarios that invoke physics beyond the Standard Model, such as cosmic strings and phase
transitions in the early Universe, which would provide exotic sources of gravitatonal waves.

The good news is that many of the puzzles and questions raised by the PTA and JWST data may soon be answered by new instalments of data from
these instruments, as well as data from other detectors that are operating or planned. Gravity research has never been more exciting.

\section*{Acknowledgements}
This work was supported by the European Regional Development Fund through the CoE program grant TK133 and by the Estonian Research Council grants PSG869, PRG803 and RVTT7. The work of J.E. and M.F. was supported by the United Kingdom STFC Grants ST/T000759/1 and ST/T00679X/1. The work of V.V. was partially supported by the European Union's Horizon Europe research and innovation program under the Marie Sk\l{}odowska-Curie grant agreement No. 101065736.

\bibliographystyle{JHEP}
\bibliography{refs}

\end{document}